# FeARH: Federated machine learning with Anonymous Random Hybridization on electronic medical records

Jianfei Cui[1], He Zhu[2], Hao Deng[3,4], Ziwei Chen[5,*], Dianbo Liu[3,4,6,7,*]
1. Viterbi School of Engineering, University of Southern California, Los Angeles, CA, 90007
2. The Hong Kong Polytechnic University, Hong kong
3. Harvard Medical School, Boston, MA 02115
4. Massachusetts General Hospital, Boston, MA 02115
5. Beijing Jiaotong University. Beijing
6. The Broad institute of MIT and Harvard, Cambridge, MA 02115
7. Computer Science & Artificial Intelligence Laboratory, Massachusetts Institute of Technology, Cambridge, MA, USA, 02139

*. Corresponds to: dianbo@mit.edu or dianbo@broadinstitute.org or zwchen@bjtu.edu.cn

# Abstract

Electrical medical records are restricted and difficult to centralize for machine learning model training due to privacy and regulatory issues. One solution is to train models in a distributed manner that involves many parties in the process. However, sometimes certain parties are not trustable, and in this project, we aim to propose an alternative method to traditional federated learning with central analyzer in order to conduct training in a situation without a trustable central analyzer. The proposed algorithm is called "federated machine learning with anonymous random hybridization (abbreviated as 'FeARH')", using mainly hybridization algorithm to degenerate the integration of connections between medical record data and models' parameters by adding randomization into the parameter sets shared to other parties. Based on our experiment, our new algorithm has similar AUCROC and AUCPR results compared with machine learning in a centralized manner and original federated machine learning.

# keywords

medical record, privacy, federated machine learning, hybridization

# 1. Introduction

Data accessing is one of the obstacles for making full use of Electronic Health Record (EHR) data, which is very valuable for precision medicine and the improvement for medical care [1,6,8,20,21]. Those medical records, generated in medical devices and doctors' diagnoses, are stored in different locations and data silos, especially in hospitals [2,10,16]. Distributed health records are usually centralized into a database and then be accessed for analysis [7,9]. As a result of regulation and privacy preservation, medical information transfer is very complex, which not only leads to high data utilization costs, but also slows down the flow of information in healthcare, as timely updates are often important [13].

Traditionally, supervised machine learning for data analysis contains two main parts that are model training, in which some datasets are used to optimize the learning model parameters, and prediction, in which the trained model uses unseen data to make predictions [3]. Federated machine learning is to train models in a distributed manner with data from different data silos that models are sent to different data sources for distributed training and then distributed-trained models are sent back to analyzer to get a federated model [4,14]. Under a distributed manner of machine learning, models can be built on data sets that are distributed across multiple devices without centralizing the data and the risk of data leakage is going to be reduced. [15,18].

However, in federated learning, many parties, such as data owners and analyzer, are involved in the whole training procedure among which data or models' parameter transfer and exchange are needed [12,15], but sometimes certain parties are not trustable, as for the federated learning process for medical data, central analyzers not trustable sometimes [4,18,19]. Different techniques have been proposed to preserve privacy when sharing information among parties in the distributed learning paradigm, such as Secure Multi-party Computation (SMC), Differential Privacy and Homomorphic Encryption [18]. But SMC guarantees complete zero knowledge that requires complicated computation protocols and may not be achieved efficiently, and Differential Privacy adds noise to data while still requiring transmitting data elsewhere. Compared with SMC and Differential Privacy, Homomorphic encryption only exchanges parameters under the encryption mechanism during machine learning that can be achieved relatively efficiently and requires no transmission of the data. Previous work has already been explored in this direction of avoiding sharing raw data. GLORE and EXPLORE have tried to share secured intermediate information [18, 22, 23], while compared to our work, they have focused on logistic regression models but we focus on Neural networks, and we have added randomization parameters into the shared neural-network model. Aono has tried scalable and secure logistic regression via homomorphic encryption in 2016, but his work didn't focus on neural networks neither [24], and most of previous works have followed a predefined protocol. Therefore, we have tried to implement hybridization in parameters exchange to break down the integration of parameter sets and add randomization into the shared parameters without strictly following a predefined protocol and we will focus on neural network models' parameters, as a result that we also have proved the effectiveness in Logistic regression model (The result is posted in Table 1_2), but the neural network model experiments demonstrated better performance, and we think that our proposed algorithm strategy can be useful for multiple machine learning models that adopting batches of parameters but the analysis or experiments of other models is beyond this paper's scope.

We proposed a new method, Federated machine learning with Anonymous Random Hybridization (FeARH), to deal with privacy problems derived from untrustable central analyzer in the federated machine learning process in a distributed manner. More specifically, the proposed method is to deal with "active" adversary which means that the central untrusted party can deviate from the protocol and maliciously modify model parameters in order to obtain sensitive information about the training sets. With a hybridization algorithm, parties of locally training exchange parts of their models' parameters with other parties before all being delivered to other institutions as central analyzer and the action of distribution can be initialized by either any of the data owners or a third party analyzer who can only send the initial models' parameters or partial and randomized parameter sets, under this strategy, not only the medical data won't be shared to other institutions but also the data owners won't share an integrated parameter set with others, which is replaced by a randomized and partial set of parameters,  so the participants as data owners and analyzers will only receive partial or randomized models that make them unlikely to backtrack on data used for training, then medical information that belongs to local data owners will less likely be leaked and privacy will better preserved [17]. Compared with the original federated learning, the main difference of FeARH is shown in figure 1 as below.

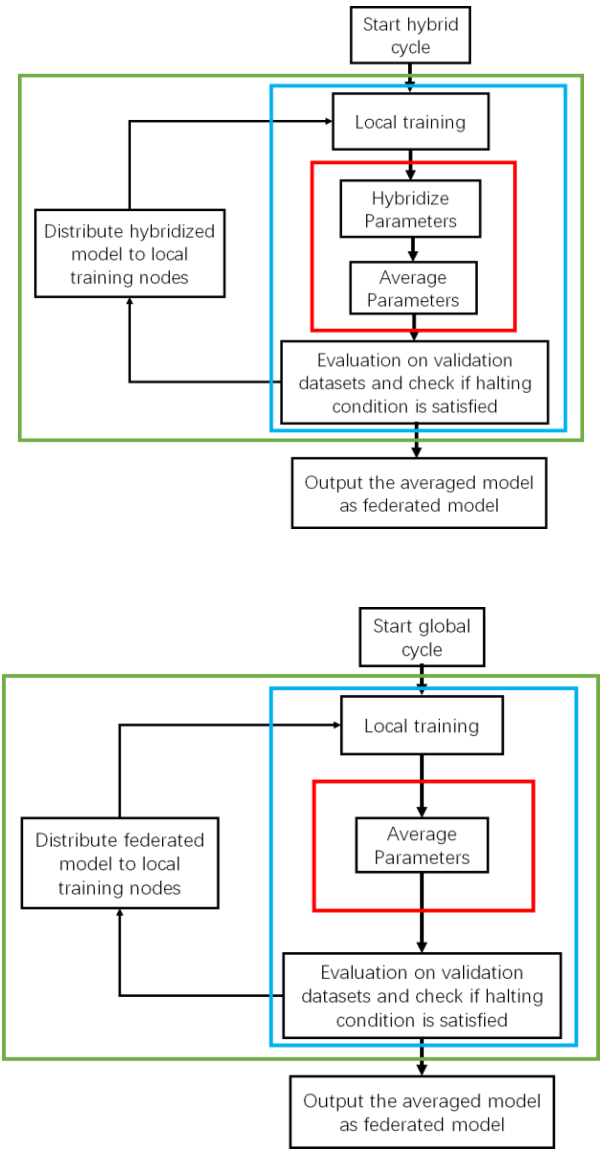

Figure 1: Brief flowchart of FeAEH (up) vs original federated learning (down), the most important difference is the existence of hybridization that is marked in red rectangle, and the local cycle is in blue rectangle, and hybrid cycle (called global cycle in original federated learning) is in green.

# 2. Method

## 2.1 Dataset

The eICU Collaborative Research Database is a large multi-center critical care database made available by Philips Healthcare in partnership with the MIT Laboratory for Computational Physiology [25]. We include data of 30'760 patients' information about the medicines taken by each patient and mortality of each patient. We have developed neural network models that take the medicines taken-in as input features to predict mortality, and we have 2913 features. Binary information of whether a patient took each of the medicines after admission is used to describe input features, and the class label is also represented by binary that 1 represents patients' death. There are 30.5% of patients recorded as being dead in our dataset, while others are recorded as being alive.

## 2.2 Model

In order to make predictions in binary for each patient based on their medicine taken situation, we built a 3-layer fully connected artificial neural network model, used for local training with 4, 2, and 1 neurons in corresponding layers using 'Nadam' as the optimizer while the weights initializer is 'glorot_uniform' and the bias initializer is 'zeros'. And we use cross-entropy as loss function that is:

$$argmin - \sum_{\omega=1}^{N} \quad [y_\omega log f(x_\omega) + (1 - y_\omega) log(1 - f(x_\omega))] \tag{1}$$

Where $x_\omega$ is the feature vector with dimension (1,2913), $y_\omega$ is the binary mortality label that 0 for alive and 1 for dead, $f$ is the neural network model and $N$ is the total number of patients used for training.

## 2.3 Federated learning setting

We have 30'760 patients' information about their medicine taken and mortality, and try to train a model to predict mortality in a distributed manner given taken-in medicines. We split all the data that has been shuffled into three parts: a training dataset containing 70% of all data, a testing dataset containing 20% of all data and a validation dataset containing 10% of all data. In each stage of our experiments (train, validation, test) to mimic the real world medical setting, we assume that we have $n$ data owners who evenly share the data points, and we set $n = 8$ in our experiment that we have 3*8=24 sub-datasets $\{T_1, T_2, \ldots, T_n\}$, $\{E_1, E_2, \ldots, E_n\}$ and $\{V_1, V_2, \ldots, V_n\}$, representing local data used for training, testing and validation in a distributed manner as shown in figure 2.

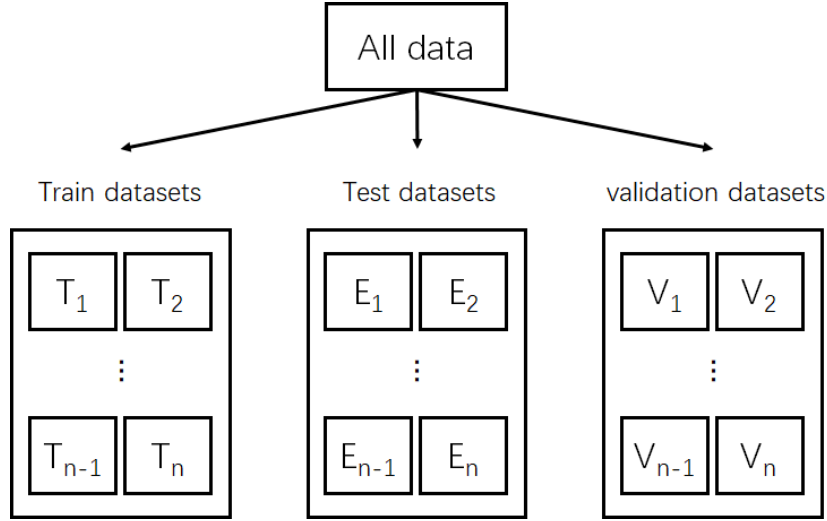

Figure 2: Structure of data overall split for training, validation and evaluation

## 2.4 federated machine learning with anonymous random hybridization

In our project, the main goal is to deal with untrustable central analyzers in federated learning. And details of simulation of the proposed algorithm, FeARH, are illustrated as below.

In order to implement this situation in program where we use python for programming and call Keras (2.2.4 version and tensorflow backed) to import machine learning models, we set $n$ neural network models $\{m_1, m_2, \ldots, m_n\}$ as described in Section 2.2 and we distribute all models to each data owner one by one. In each node, the neural network model $m_i: i \in [1, n]$ is trained by gradient descent, after all $n$ training processes finished, we randomly divide all $n$ models in pairs, then let these 2 models of each pair hybridize their parameters. In each neural network model, we count that there are $\lambda = 11'669$ parameters, denoted as $\{p_1, p_2, p_3, \ldots, p_\lambda\}$. With hybrid exchange rate $\gamma$, there are $\gamma \times \lambda$ random-selected parameters to be exchanged between two models, and for one certain parameter, it's changed to the same position in the other model.

**Hybrid Algorithm**

1: **def** Hybrid $(m, n, \gamma)$
2:     $i \leftarrow 0$
3:     model_index=random.sample(range(n), int(n/2))
4:     **for** $i \leqslant n/2$ **do**
5:         flat($m_{index[2i]}, m_{index[2i+1]}$) #reshape parameters into a 1-D
6:         oneZeroArray←numpy.zeros($m_{index[2i]}$.shape)
7:         bit_index←random.sample(range($m_{index[2i]}$.shape[1]),
                        int($m_{index[2i]}$.shape[1]$* \gamma$))
                     # select the position of parameters
                       to be exchanged to other models
8:         $j \leftarrow 0$
9:         **for** $j \leqslant m_{index[2i]}.shape[1] * \gamma$ **do**
10:           oneZeroArray [index[$j$]]← 1
11：       $m_{index[2i]} \leftarrow m_{index[2i+1]} *$oneZeroArray$+m_{index[2i]} *$(-(oneZeroArray-1))
12：       $m_{index[2i+1]} \leftarrow m_{index[2i]} *$oneZeroArray$+m_{index[2i+1]} *$(-(oneZeroArray-1))
14:     **return** $m$

After hybrid process, all $n$ models $\{m_1, m_2, \ldots, m_n\}$ is partly changed that there are $\gamma \times \lambda$ parameters belonging to another model before hybridization, so the integration of parameter sets from local training dataset owners is degenerated as shown in figure 3.

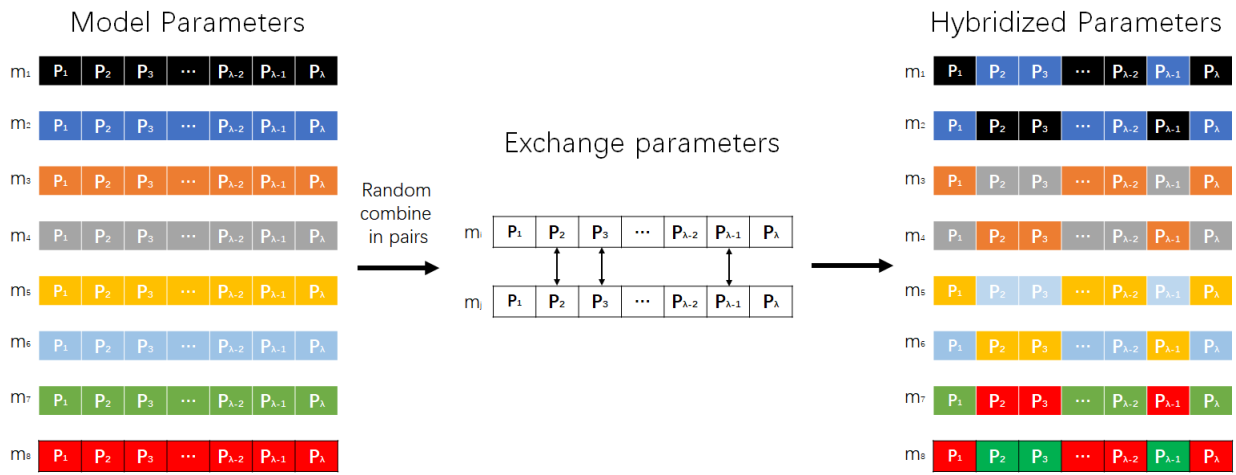

Figure 3: The behavior of parameters in hybridization algorithm

After the training process in all data owners, the analyzer will receive all parameters of $n$ models and the percent of the number of data pieces used for training each model, which are denoted as $\{\rho_1, \rho_2, \rho_3, \ldots, \rho_n\}$ and $\sum_{l=1}^{n} \rho_l = 1$, then the analyzer will average these models' hybridized parameters based on the percent of data pieces for training each one and the average algorithm and equation (2).

**Average algorithm**

1: **def** average $(\rho, m)$
2:    $M_{fed} \leftarrow 0, i \leftarrow 0$
3:    **for** $i \leqslant n$ **do**
4:        $M_{fed} \leftarrow m_i * \rho_i$
5:        $i + 1$
6: **return** $M_{fed}$

$$M_{fed} = \sum_{l=1}^{n} \rho_l \times m_l \qquad (2)$$

After local training stage, $M_{fed}$ will be distributed to validation datasets $\{V_1, V_2, \ldots, V_n\}$ and calculate the average AUCROC score on all $n$ validation datasets. If the averaged AUCROC score does not increase by at leaset 0.01% for three successive times, we will stop the training and validation stage (one hybrid cycle contains on training stage and one validation stage) , otherwise, we will distribute hybridized model $\{m_1, m_2, \ldots, m_n\}$ to $\{T_1, T_2, \ldots, T_n\}$ correspondingly and repeat the hybrid cycle.

After hybrid cycle, we get federated model's parameter $M_{fed}$, we will distribute $M_{fed}$ to validation datasets $\{V_1, V_2, \ldots, V_n\}$ and calculate the average AUCROC score of all $n$ validation datasets. If the averaged AUCROC score does not increase by at leaset 0.01% for three successive times, we will stop the training and validation stage (one hybrid cycle contains one training stage and one validation stage) , otherwise, we will re-distribute $\{m_1, m_2, \ldots, m_n\}$ (Hybridized parameters) to $\{T_1, T_2, \ldots, T_n\}$ correspondingly and repeat the hybrid cycles.

After hybrid cycles, we will distribute the federated model at saturation point, $M_{fed}$, that after which the average AUCROC score doesn't increase by at least 0.01%, to $\{E_1, E_2, \ldots, E_n\}$ and conducting testing in a distributed manner. AUCROC score and AUCPR score will be calculated for final evaluation by averaging locally AUCROC score and AUCPR score at each silo containing testing dataset.

Moreover, privacy preservation has been a prominent concern for collaborative learning. Therefore, we have compared the performance of both the FeARH model and the original federated learning model in protecting sensitive information from leakage. The detailed information has been demonstrated in the Appendix section.

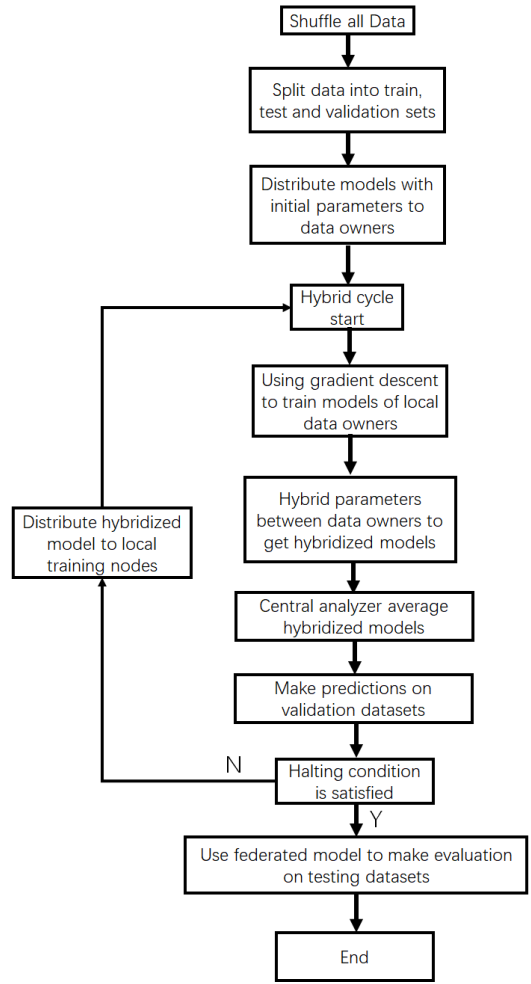

Figure 4: Flowchart of simulation of the new proposed algorithm 'federated machine learning with anonymous random hybridization (FeARH)'

Table 0: Notation

| Notation | Explanation |
|---|---|
| $n$ | The number of local data owners |
| $T$ | Datasets used for training |
| $E$ | Datasets used for testing |
| $V$ | Datasets used for validation |
| $m_i$ | The $i\,th$ models or parameter set |
| $\lambda$ | Total number of parameters in one model |
| $p_i$ | The $i\,th$ parameter in one model |
| $\gamma$ | Exchange rate |
| $\varrho_i$ | The $i\,th$ data owner's' percentage of data over total dataset size |
| $M_{fed}$ | Federated model or parameter set |

# 3. Result and Discussion

## 3.1 AUCROC and AUCPR

To evaluate the proposed algorithm, Federated machine learning with Anonymous Random Hybridization (FeARH), we have conducted an experiment to compare the AUCROC score and AUCPR score of centralized learning, original federated learning and FeARH, and to make them comparable to each other, all three methods apply to the same halting condition that if the AUCROC score on validation datasets doesn't increase by at least 0.01%, the experiments' training part will stop and move to making predictions on test datasets for evaluation.

Due to the medical mortality information from eICU database we used is imbalanced, we use both AUCROC and AUCPR to evaluate our model's accuracy [5,6].

When we adopted a centralized manner to train the neural network model as described in section 2.2, the model achieved AUCROC of 0.8412 and AUCPR of 0.6794 on the test data (Table1).

Then we conducted our experiment using original federated learning and the main difference compared with FeARH is demonstrated in figure 1. In original paradigm, neural network models $\{m_1, m_2, \ldots, m_n\}$ that each is the same as what have been discussed in 2.2 are distributed to local datasets $\{T_1, T_2, \ldots, T_n\}$ for local training with gradient descent. Then the trained models are sent directly to the analyzer without a hybrid process, and the analyzer will use average algorithm to assemble locally trained models into one federated neural network model $M_{fed}$, which will be used for validation with $\{V_1, V_2, \ldots, V_n\}$. Then we have the same halting condition as FeARH that if the averaged AUCROC score on validation datasets doesn't increase by at least 0.01%, the global cycle that contains one local training and one validation will stop. Otherwise we will distribute federated model $M_{fed}$ back to $\{T_1, T_2, \ldots, T_n\}$ and repeat the global cycle until the halting condition is satisfied. After the global cycle, we use testing datasets $\{E_1, E_2, \ldots, E_n\}$ for evaluation in a distributed manner. The federated model's performance of original federated learning achieved AUCROC of 0.8452 and AUCPR-score of 0.7019 (Table 1).

Next, we trained the model using a FeARH strategy, all our local models are trained in distributed manner using gradient descent with $\{T_1, T_2, \ldots, T_n\}$, after which we hybridize our models and let them exchange parameters. After hybridization, average algorithm is executed for federation and the federated model will be sent to $\{V_1, V_2, \ldots, V_n\}$ for validation. If the halting condition is satisfied, the hybrid cycle will stop and the algorithm moves into making predictions on test datasets $\{E_1, E_2, \ldots, E_n\}$ for evaluation in a distributed manner, and we achieve an averaged AUCROC of 0.8312 and AUCPR of 0.6778 as shown in table 1.

Table 1_1:  Performance of centralized, original federated and FeARH with Neural Network

| Method | AUCROC | AUCPR |
|---|---|---|
| Centralized Learning | 0.8412 | 0.6794 |
| Original federated machine learning | 0.8452 | 0.7019 |
| Federated machine learning with anonymous random hybridization (FeARH) | 0.8312 | 0.6778 |

Table 1_2: Performance of centralized, original federated and FeARH with logistic regression

| Method | AUCROC | AUCPR |
|---|---|---|
| Centralized Learning | 0.8412 | 0.4579 |
| Original federated machine learning | 0.7816 | 0.5989 |
| Federated machine learning with anonymous random hybridization (FeARH) | 0.7864 | 0.6206 |

## 3.2 AUCROC, AUCPR  VS  Hybrid exchange rate

We also have explored the impact of hybrid exchange rate $\gamma$ on the model's performance. In the figure 5, we compare AUCROC, AUCPR with different exchange rate $\gamma$, We set hybrid exchange rate $\gamma$ at 0.1~0.5 with an increment of 0.1. In figure 5, the numbers in horizontal axis represent the number of hybrid exchange rate $\gamma$, and we can see from figure 5 that both AUCROC and AUCPR will decrease as exchange rate $\gamma$ increasing in the given range. Generally speaking, higher exchange rates lower the risk of leakage of private data, but in practice, the level of privacy is a subjective choice in specific cases, and the actual value of $\gamma$ to use depends on real world application.

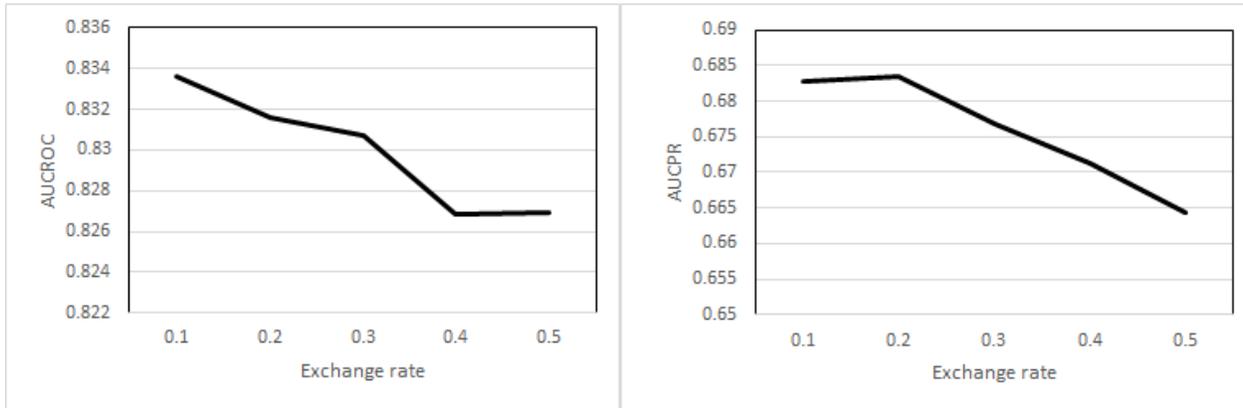

Figure 5: The exchange of AUCROC (left) and AUCPR (right) performance as the change of hybrid exchange rate

## 3.3 Number of hybrid cycles

In real world setting, the number of hybrid cycles is determined by the evaluation on validation datasets, in our simulation experiments of FeARH, we get an average number of 5.6 cycles before the halting condition is satisfied when conducting 50 times of our experiments with a exchange rate of 0.1 (this is a subjective parameter as described in Section 3.2). At the same time, we also have conducted original federated learning for 50 times and reach an average number of global cycles of 5.2.

---

**Federated machine learning with Anonymous Random hybridization (FeARH)**

1:**Procedure** FeARH$(n, \rho)$
2:　　initial models with same weights and bias

```
3:      while halting condition is False do
4:          for j in [1, n] do
5:              train models at local data owner $j$ and obtain $m_j$
6:          hybridize parameters $\{m_1, m_2, \ldots, m_n\}$
7:          average parameters by $M_{fed} = \sum_{l=1}^{n} \rho_l \times m_l$
8:          Calculate AUCROC on validation datasets
9:      return $M_{fed}$
```

# 4. Conclusion

We proposed a distributed machine learning training method, federated machine learning with anonymous random hybridization (FeARH), that aims to deal with untrustable parties in federated machine learning by implementing a hybrid algorithm with adding randomization into parameters sets shared among different parties. We also conducted experiments to evaluate our proposed algorithm and showed that FeARH had similar performance compared with machine learning in a centralized manner and original federated learning paradigm.

# 5. Appendix

Furthermore, we have conducted another experiment to evaluate the performance of both the FeARH model and the conventional federated learning model in terms of privacy preservation. Specifically, we have constructed and trained a Generative Adversarial Network (GAN) as our attack model, which generates and reconstructs the sensitive information by influencing the training process and deceiving the victims to reveal more detailed information [26]. We assume that the attacker has the prior knowledge of the model structure and, in particular, of the data labels of other participants in the collaborative network. Besides, we also assume that the victim declares the labels ["0", "1"], which represent the binary mortality information, while the attacker declares the labels ["0", "2"], where the label "2" denotes the unknown information. In order to infer the meaningful information about a label he does not own, the attacker creates a replica of the freshly updated local model as the discriminator and then trains a local generative adversarial network to mimic the samples with label "1" from the victim. Then he deliberately assigns a fake label "2" to those samples, which surreptitiously obfuscates the model parameters during the collaborative training process. In this case, the victim needs to work harder to distinguish between the samples labelled with "1" and "2", and thus reveal more information about label "1". Moreover, we have also trained a solitary classifier to distinguish the implausible samples generated by the attacker between the authentic samples with label "1", so as to evaluate the performance of the attacker. The result demonstrated in Table 2 shows that the attacker generally performs worse on the FeARH model than the conventional federated learning model, which implies that our proposed algorithm is more resistant to information leakage, and thus more privacy preservation than the conventional federated learning paradigm.

Table 2: Performance of the attacker on the original federated and FeARH model

| Method | Accuracy | Loss |
| --- | --- | --- |
| Original federated machine learning | 0.9957 | 0.0976 |
| Federated machine learning with anonymous random hybridization (FeARH) | 0.6919 | 3.0011 |

# Acknowledgement and funding

JC conceptualized the project, designed the study, conducted all the analysis and experiments. ZH conducted the experiment to evaluate the performance of the proposed algorithm. DL conceptualized the project, designed the study, designed the experiment and supervised the project. ZWC contributed to algorithm design. Part of ZWC's work on this project was conducted at MIT. HD contributed to medical experiments design.

ZWC was under the financial support of the "the Fundamental Research Funds for the Central Universities" (2019RC032).

Other authors were not under any specific funding support